\documentclass[ms]{egu} % (for Atmos. Chem. Phys.) please find the list of
\printfigures

\begin{document}
\title{A simple conceptual model of abrupt glacial climate events.}
\runningtitle{A conceptual model of Dansgaard-Oeschger events}  % abbreviated title
\runningauthor{H.~Braun et al.} % "First and Second" or "First et al."
\correspondence{H.~Braun\\ (Holger.Braun@iup.uni-heidelberg.de)} % with email
% VARIOUS WAYS TO ENTER THE AUTHORS' NAMES AND AFFILIATIONS
%
%   1. IF ALL AUTHORS ARE FROM THE SAME INSTITUTE:
%      Enter authors with separate \author commands for each author, and
%      then the affiliation with \affil.
%
%\author{J.~G.~Smith}
%\author{H.~K.~Weston}
%\affil{Institute for Historical Geophysics, Houston, Texas}
%
%   2. IF THE AUTHORS ARE FROM SEVERAL INSTITUTES:
%      Enter all authors from the first institute with separate \author
%      commands for each author, and then the affiliation for
%      the above authors with \affil. Repeat for the next institute.
%      This method requires that the authors are sorted by institutes.
%      Affiliation superscripts are added automatically.

\author{H.~Braun}
\affil{Heidelberg Academy of Sciences and Humanities, c/o Institute of
  Environmental Physics, University of Heidelberg, Im Neuenheimer Feld 229,
  69120 Heidelberg, Germany.} 
\author{A.~Ganopolski}
\affil{Potsdam Institute for Climate Impact Research, P.O. Box 601203, 14412
  Potsdam, Germany.}  
\author{M.~Christl}
\affil{PSI/ETH Laboratory for Ion Beam Physics, c/o Institute of Particle
  Physics, ETH Zurich, 8093 Zurich, Switzerland.} 
\author{D.~R.~Chialvo}
\affil{Department of Physiology, Feinberg Medical School, Northwestern
  University, 303 East Chicago Ave. Chicago, IL 60611, USA.}
%\affil{Institute for Historical Geophysics, Houston, Texas}

%\author{T.~P.~Phillips}

%\affil{School for Military Advances, London, United Kingdom}

%

%   3. IF SPECIAL SUPERSCRIPTS ARE REQUIRED:

%      Enter all authors in the desired order, then all institutes.

%      Affiliation superscripts (and further footnotes) must be entered

%      explicitly in square brackets.

%

%\author[1]{J.~G.~Smith}

%\author[2]{T.~P.~Phillips}

%\author[1,*]{H.~K.~Weston}

%\affil[1]{Institute for Historical Geophysics, Houston, Texas}

%\affil[2]{School for Military Advances, London, United Kingdom}

%\affil[*]{Present address Playa del Ingeles, Gran Canaria}

\maketitle % YOU MUST USE THE \maketitle COMMAND

\begin{abstract}
Here we use a very simple conceptual model in an attempt to reduce essential
parts of the complex nonlinearity of abrupt glacial climate changes (the
so-called Dansgaard-Oeschger events) to a few simple principles, namely (i)
the existence of two different climate states, (ii) a threshold process and
(iii) an overshooting in the stability of the system at the start and the end
of the events, which is followed by a millennial-scale relaxation.
By comparison with a so-called Earth
system model of intermediate complexity (CLIMBER-2), in which the events represent
oscillations between two climate states corresponding to two fundamentally
different modes of deep-water formation in the North Atlantic, we demonstrate
that the conceptual model captures fundamental aspects of the nonlinearity of
the events in that model. We use the conceptual model in order to reproduce
and reanalyse nonlinear resonance mechanisms that were already suggested in
order to explain the characteristic time scale of Dansgaard-Oeschger
events. In doing so we identify a new form of stochastic resonance (i.e. an
{\it{overshooting stochastic resonance}}) and provide the first explicitly
reported manifestation of {\it{ghost resonance}} in a geosystem, i.e. of a
mechanism which could be relevant for other systems with thresholds and with
multiple states of operation. Our work enables us to explicitly simulate
realistic probability measures of Dansgaard-Oeschger events (e.g. waiting
time distributions, which are a prerequisite for statistical analyses
on the regularity of the events by means of Monte-Carlo
simulations). We thus think that our study is an important advance in
order to develop more adequate methods to test the statistical
significance and the origin of the proposed glacial 1470-year climate cycle.   
\end{abstract}

\introduction\label{sec:intro} 

\begin{figure}[t]
\vspace*{2mm}
  % don't type the extension (eps or pdf) of the graphics file here
  \begin{center}
  \includegraphics[width=8cm]{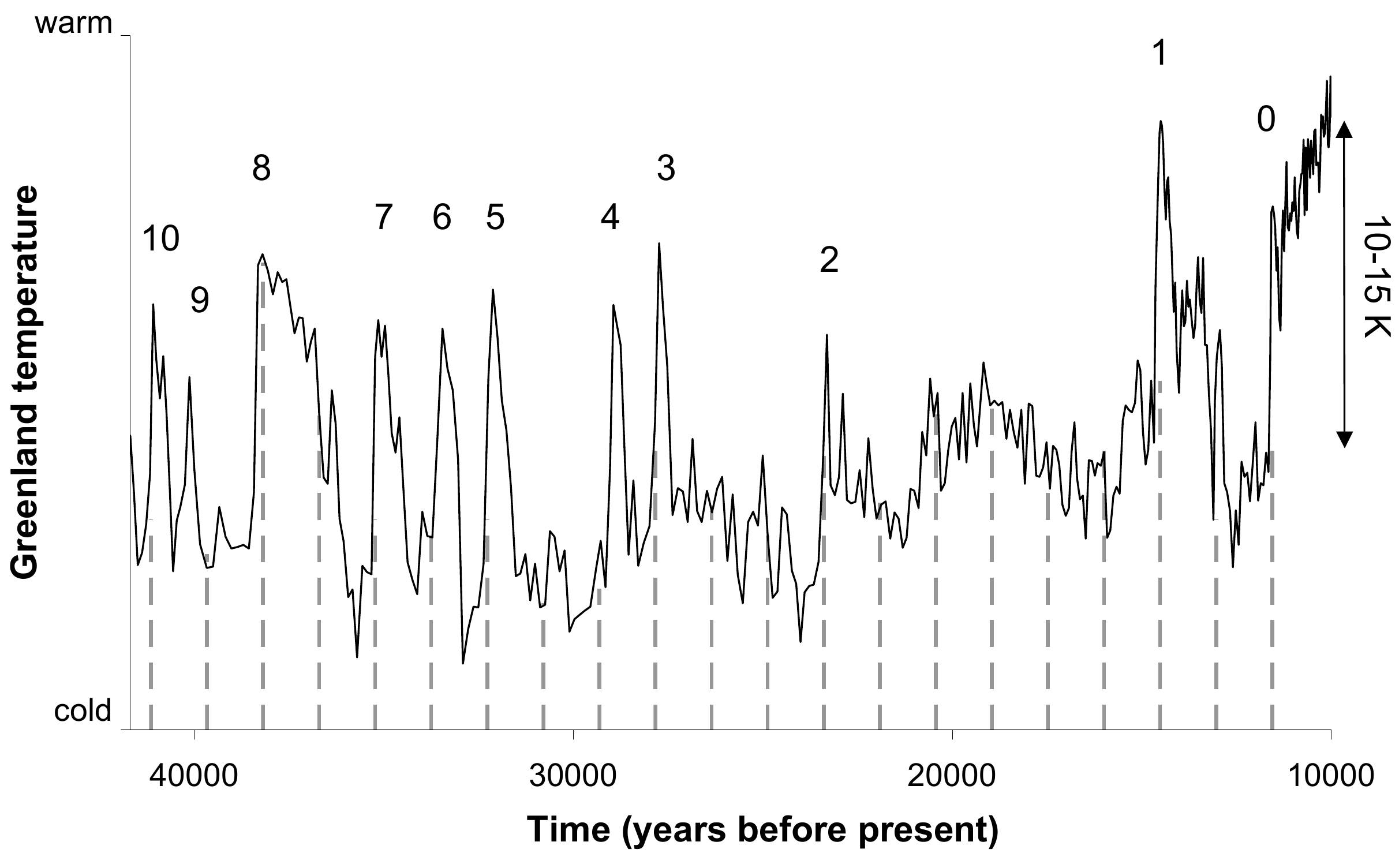}
  \end{center}
  \caption{\label{fig:figure1}
DO events as seen in the GISP2 ice-core from Greenland \citep{Grootes1993, Grootes1997}. The figure shows Greenland
temperature changes over the interval between 10000 and about 40000 years
before present. DO events (0-10) manifest themselves as saw-tooth shaped warm
intervals.  Dashed lines are spaced by 1470 years.
}
\end{figure}

Time series of North Atlantic atmospheric / sea surface temperatures during the last ice age
reveal the existence of repeated large-scale warming events, the so-called Dansgaard-Oeschger 
(DO) events \citep{Dansgaard1982, Grootes1993}.
In climate records from the North Atlantic region the events have a
characteristic saw-tooth shape (Fig.~\ref{fig:figure1}): They typically
start with a warming by up to 10-15 K \citep{Severinghaus1999,
  Leuenberger1999} over only a few years/decades. Temperatures remain high for
centuries/millennia until they drop back to pre-events values over a century
or so. A prominent feature of DO events is their millennial time scale: During
Marine Isotope Stages (MIS) 2 and 3, successive events in the GISP2 ice core
were reported to be often spaced by about 1470 years or multiples thereof
\citep{Alley2001a, Schulz2002, Rahmstorf2003}, compare
Fig.~\ref{fig:figure1}. A leading spectral peak corresponding to the 1470-year
period was found \citep{Grootes1997, Yiou1997}, and this spectral component was
reported to be significant at least over a certain time interval \citep{Schulz2002}.
We note, however, that the statistical significance of this pattern is still under
debate \citep{Ditlevsen2007}, in particular because of the lack of adequate
nonlinear analysis methods.  

It was proposed that DO events represent rapid transitions between two fundamentally different 
modes of the thermohaline ocean circulation (THC) \citep{Oeschger1984, Broecker1985}, most 
likely corresponding to different modes of deep-water formation \citep{AlleyClark1999, 
Ganopolski2001}. The origin of these transitions is also under debate: 
In principle they could have been caused by factors from outside of the Earth
system \citep{Keeling2000, Rial2004, Clemens2005, Braun2005}, but they
could also represent internal oscillations of the climate system
\citep{Broecker1990, Sakai1997, Kreveld2000}.  
Several nonlinear resonance mechanisms have been suggested in order 
to explain the characteristic timing of DO events, including coherence
resonance \citep{Ganopolski2002, Timmermann2003, 
Ditlevsen2005} and stochastic resonance \citep{Alley2001a, Alley2001b,
Ganopolski2002, RahmstorfAlley2002}.

\section{Spectrum of models}
\label{sec:spectrum}

DO events have already been simulated by a variety of models, ranging from simple conceptual 
ones to Earth system models of intermediate complexity (EMICs). Conceptual models are most 
suitable to perform large numbers of long-term investigations because they require very
little computational cost. However, they are often based on ad-hoc assumptions and only 
consider processes in a highly simplified way. In addition to that, the number of adjustable 
parameters is typically large compared to the degrees of freedom in those models. This implies 
that seemingly good results can often be obtained merely by excessive tuning. Nevertheless, 
conceptual models can provide important help for the interpretation of complex climatic 
processes. 

The gap between conceptual models and the most comprehensive general circulation 
models (GCMs), which are not yet applicable for millennial-scale simulations because of their 
large computational cost, is bridged by EMICs \citep{Claussen2002}. EMICs include 
most of the processes described in comprehensive models (in a more reduced form), and 
interactions between different components of the Earth system (atmosphere, hydrosphere, 
cryosphere, biosphere, etc.) are simulated. The number of degrees of freedom typically 
exceeds the number of adjustable parameters by orders of magnitude. Since many EMICs are fast
enough for studies on the multi-millennial time scale, they are the most adequate tools for the 
simulation of DO events.  

\section{The conceptual model}
\label{model}

The simple conceptual model which we use here is an extended version of the model
described by \citet{Braun2005} (in the Supplementary Material of that publication).
Here we use the model to demonstrate and analyse two apparently counterintuitive
resonance phenomena ({\it{stochastic resonance}} and {\it{ghost resonance}})
that can exist in a large class of highly nonlinear systems. Due to the
complexity of many of those systems it is often impossible to precisely
identify the reasons for the occurrence of these resonance phenomena. Our
conceptual model, in contrast, has a very clear forcing-response relation as
well as a very low computational cost and thus provides a powerful tool to
explore these phenomena and to test their robustness. Furthermore, we describe
and discuss the applicability of the model for improved statistical analyses
(i.e. Monte-Carlo simulations) on the regularity of DO events. In the
following the key assumptions of the conceptual model are first discussed. In
the Supplementary Material we then compare the model performance under a
number of systematic forcing scenarios with the performance of a more
comprehensive model (the EMIC CLIMBER-2), compare Supplementary Information
File and Supplementary Figs. 1-6. In the framework of the conceptual model we
finally demonstrate and interpret two hypotheses that were previously
suggested in order to explain the recurrence time of DO events, and we discuss
how these hypotheses could be tested.       

\subsection{Model description}

Our conceptual model is based on three key assumptions: 
\begin{enumerate}
\item DO events represent repeated transitions between two different climate states, 
corresponding to warm and cold conditions in the North Atlantic region. 
\item These transitions are rapid compared to the characteristic life-time of
  the two climate states (i.e. in first order approximation they occur
  instantaneously) and take place each time a certain threshold is crossed. 
\item With every transition between the two states the threshold overshoots
  and afterwards approaches equilibrium following a millennial-scale
  relaxation process. This implies that the conditions for a switch between
  both states ameliorate with increasing duration of the cold and warm
  intervals. 
\end{enumerate}

Our three assumptions are supported by paleoclimatic evidence and/or by
simulations with a climate model:
\begin{enumerate}
\item Since long, DO events have been regarded as repeated oscillations between two 
different climate states \citep{Dansgaard1982}. It has been suggested that
these states are linked with different modes of operation of the THC
\citep{Oeschger1984, Broecker1985}. This seminal interpretation has since then
influenced numerous studies and is now generally accepted \citep{Rahmstorf2002}. 
Indirect data indicate that   
the glacial THC indeed switched between different modes of operation \citep{Sarnthein1994, 
AlleyClark1999} which, according to their occurrence during cold and warm 
intervals in the North Atlantic, were labelled stadial and interstadial modes. A third 
mode named Heinrich mode (because of its presence during the so-called Heinrich events) 
is not relevant here.
\item High-resolution paleoclimatic data show that transitions from cold conditions in 
the North Atlantic region to warm ones often happened very quickly, i.e. on the 
decadal-scale or even faster \citep{Taylor1997, Severinghaus1999}. The opposite
transitions were slower, i.e. on the century-scale \citep{Schulz2002}, but
nevertheless still faster that the characteristic life-time of the cold and
warm intervals (which is on the centennial to multi-millennial time scale, compare Fig. 1). The 
abruptness of the shifts from cold conditions to warm ones has commonly been
interpreted as evidence for the existence of a critical threshold in the
climate system that needs to be crossed in order to trigger a shift between 
stadial and interstadial conditions \citep{Alley2003}. Such a threshold could be 
provided by the THC (more precisely, by the process of deep-water formation): When warm 
and salty surface water from lower latitudes cools on its way to the North Atlantic, 
its density increases. If the density increase is large enough (i.e. if the surface gets 
denser than the deeper ocean water), surface water starts to sink. Otherwise,
surface water can freeze instead of sinking. The onset of deep-water formation can
thus hinder sea-ice formation and facilitate sea-ice melting (due to the 
vertical heat transfer between the surface and the deeper ocean). A switch between two 
fundamentally different modes of deep-water formation can thus dramatically change sea 
ice cover and can cause large-scale climate shifts. Such nonlinear, threshold-like 
transitions between different modes of deep-water formation are at present considered as 
the most likely explanation for DO events \citep{Alleyetal1999, Ganopolski2001}.
\item The time-evolution of Greenland temperature during the warm phase of DO
  events has a characteristic saw-tooth shape
  (Fig.~\ref{fig:figure1}). Highest temperatures typically occur during the
  first decades of the events. These temperature maxima are followed by a
  gradual cooling trend over centuries/millennia, before the system returns to
  cold conditions at the end of the events. This asymmetry supports the idea
  that the system overshoots in some way during the abrupt warming at the
  beginning of the events and that the subsequent cooling trend
  represents a millennial-scale relaxation towards a new equilibrium
  \citep{Schulzetal2002,Centurelli2006}. We note that the time-evolution of
  Greenland temperature provides no clear evidence for an overshooting during
  the opposite transitions (i.e. from the warm state back to the cold
  one). This, however, is not necessarily in contradiction to our assumption:
  This lack of an overshooting in the temperature fields does not necessarily
  mean that the ocean-atmosphere system did not overshoot, since Greenland
  temperature evolution in the stadial state might have been dominated by
  factors other than the THC (respectively its stability), e.g. by Greenland
  ice accumulation, which would mask the signature of the THC in the ice core
  data.

 We will show later (in Sect. 3.3.) that the assumption of an overshooting in the
stability of the system is in fact strengthened by the analysis of model
results obtained with the coupled model CLIMBER-2. In that model the
overshooting results from the dynamics of the transitions between the two
climate states: In the stadial state deep convection occurs south
of the Greenland-Scotland ridge (i.e. at about 50 $^\circ$N). In the
interstadial state, however, deep convection takes place north of the ridge
(i.e. at about 65 $^\circ$N). The onset of deep convection north of the
Greenland-Scotland ridge, which releases accumulated energy to the atmosphere
(i.e. heat that is stored mainly in the deep ocean), in first place starts DO
events in the model. This heat release leads to a reduction of sea ice, which
in turn further enhances sea surface densities between 50 $^\circ$N and 65
$^\circ$N (e.g. by increased local evaporation and reduced sea ice transport
into that area). As a result deep convection also starts between 50 $^\circ$N
and 65 $^\circ$N, and much more heat can be released to the
atmosphere. Without a further response of the THC the system would return
quickly (within years or decades, i.e. with the convective time scale) to its
original state. In CLIMBER-2, however, the changes in deep convection trigger
a northward extension and also an intensification of the ocean circulation
(i.e. an overshooting of the Atlantic meridional overturning circulation;
compare \citet{Ganopolski2001}), which maintains the interstadial climate
state since it is accompanied by an increase in the salinity and heat flux to
the new deep convection area (at about 65 $^\circ$N). In response to the
overshooting of the overturning circulation, the system relaxes slowly (within
about 1000 years, i.e. with the advective time scale) towards the stadial
state. We note that the advective time scale corresponds to the millennial
relaxation time in our conceptual model. The model CLIMBER-2 also supports the
validity of our overshooting assumption during the opposite transition (from
the warm state back to the cold one), as we will show in Sect. 3.3. We would
like to stress that our interpretation of the processes during DO events is,
of course, not necessarily true since we can only speculate that the
underlying mechanism of the events is correctly captured by CLIMBER-2.  
\end{enumerate}

\subsection{Model formulation}

\begin{figure}[t]
\vspace*{2mm}
  % don't type the extension (eps or pdf) of the graphics file here
  \begin{center}
  \includegraphics[width=8cm]{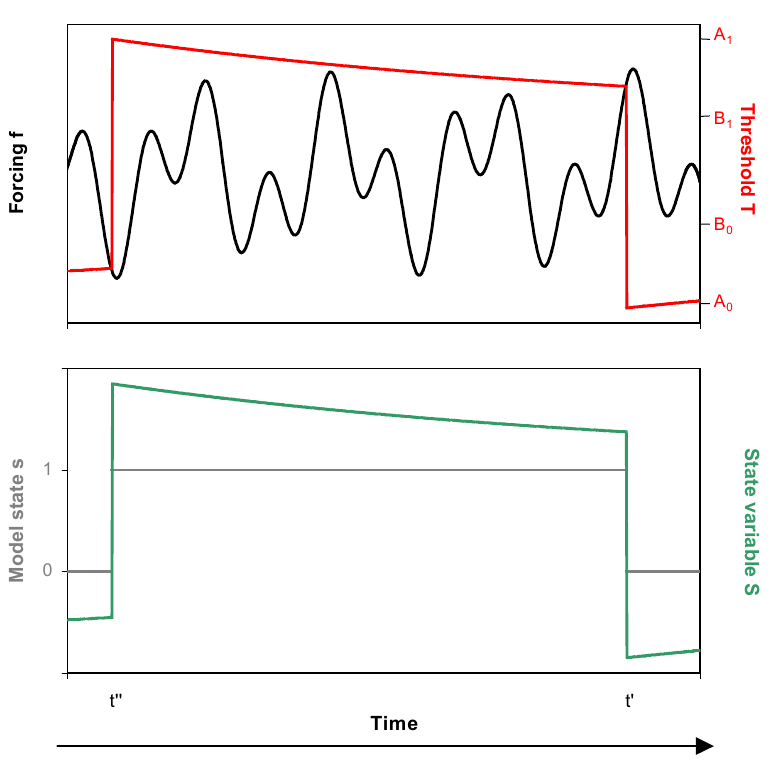}
  \end{center}
 \caption{\label{fig:figure2}
Dynamics of our conceptual model. Shown is the time evolution of the model, in response to 
a forcing that is large enough to trigger switches between both model states. Top: Forcing
f (black) and threshold function T (red). Bottom: Model state s (grey; s=0 corresponds to 
the cold state, s=1 to the warm one) and state variable S (green). At time $t''$ the forcing 
falls below the threshold function and a shift from the cold state into the warm one is 
triggered. With this transition, the threshold function switches to a non-equilibrium value 
(representing an overshooting of the system) and afterwards approaches equilibrium following 
a millennial-scale relaxation. At time $t'$ the forcing exceeds the threshold function, and a 
transition from the warm state back into the cold one is triggered. With this transition, 
the threshold function switches to another non-equilibrium value and approaches equilibrium 
following another millennial-scale relaxation, until the forcing again falls below the 
threshold function and the next switch into the warm state is triggered. Note that the 
state variable S is chosen to be identical to the threshold function T. For
convenience, discontinuities in T and S are eliminated by linear interpolation. T, S
and f are normalised in the figure. 
}
\end{figure}

We implement the above assumptions in the following way (compare
Fig.~\ref{fig:figure2}): First we define a discrete index s(t) that indicates
the state of the system at time t (in years). Since we postulate the existence
of two states, s can only take two values (s=1: warm state, s=0: cold
state). We further define a threshold function T(t) that describes the
stability of the system at time t (i.e. the stability of the current model
state).  

Second we define rules for the time evolution of the threshold function T. When the 
system shifts its state, we assume a discontinuity in the threshold function: With the switch 
from the warm state to the cold one (at time t' in Fig.~\ref{fig:figure2}) T takes the value
A$_{0}$. Likewise, with the switch from the cold state into the warm one 
(at time t'' in Fig.~\ref{fig:figure2}) T takes the value A$_{1}$. As long
as the system does
not change its state the evolution of T is assumed to be given by a relaxation process: 
\begin{equation}
\frac{dT}{dt} = - \frac{(T-B_s)}{\tau_s}
\end{equation}
(s labels the current model state, $\tau_s$ denotes the relaxation time in
that state, B$_{s}$ is 
a state-dependent constant that labels the equilibrium value of T in each model 
state). These assumptions result in the following
expression for the threshold function T: 
\begin{equation}
T(t) = (A_s-B_s) \cdot \exp(- \frac{t-\delta_s}{\tau_s})+B_s.
\end{equation} 
Note that in the above expression the index s again denotes the current state of the model
(i.e. s=0 stands for the cold state and s=1 for the warm one), 
$\delta_0$ labels the time of the 
last switch from the warm state into the cold one, and $\delta_1$ indicates the time of the last switch 
from the cold state into the warm one.

Third we assume that transitions from one state to the other are triggered each time a given 
forcing function f(t) crosses the threshold function T. More
precisely, we assume 
that when the system is in the cold state ($s[t''] = 0$) and the forcing is
smaller than the 
threshold value ($f[t''+1] < T[t''+1]$) the system switches into the warm
state ($s[t''+1] = 1$). This 
shift marks the start of a DO event. Likewise, when the system is in the warm
state 
($s[t'] = 1$) and the forcing is larger than the threshold value ($f[t'+1] >
T[t'+1]$) the system 
switches into the cold state ($s[t'+1] = 0$). That shift represents the
termination of a DO event. If none of these conditions is fulfilled, the
system remains in its present state (i.e. $s[t+1] = s[t]$).

To simplify the comparison of the model output with paleoclimatic
records we further define a state variable S, which represents anomalies in
Greenland temperature during DO events. For simplicity we assume that the
state variable is equal to the threshold function: $S(t) = T(t)$ (i.e. we
assume that Greenland temperature evolution during DO events is closely
related to the current state of the THC,
respectively to its stability). We stress that
this assumption is of course highly simplified, because Greenland
temperature is certainly not only influenced by the THC but also by other
processes such as changes in ice accumulation during DO oscillation. However,
this assumption is not crucial for the dynamics of our model, since the
timing of the switches between both model states is solely determined by the
relation between the forcing function f and the threshold function T. This
means that even if we included a more realistic relation between T and S, the
timing of the simulated climate shifts would be unchanged and the model
dynamics would thus essentially be invariant.  

Note that $\delta_0$ and $\delta_1$ are not adjustable; they rather represent
internal time markers. Thus, six adjustable 
parameters exist in our model as described here, namely $A_{0}$, $A_{1}$,
$B_{0}$, $B_{1}$, $\tau_{0}$ and $\tau_{1}$. Our choice for these parameters
is shown in Table~\ref{tab:table1}. With these parameter values the system is
bistable (i.e. no transition is ever triggered in the absence of any forcing,
since $B_{0} \le 0$ and $B_{1} \ge 0$) and almost symmetric. That
means that the average duration of the simulated warm and cold intervals is
almost equal. When compared with Greenland paleotemperature records this
situation most likely corresponds to the time interval between about 27000 and
45000 years before present, during which the duration of the cold and warm
intervals in DO oscillations was also comparable (Fig. 1). The
model can, however, also represent an
unstable (for $B_{0} > 0$ and $B_{1} < 0$) or a mono-stable system, in which the stable
state is either the warm one (for $B_{0} > 0$ and $B_{1} \ge 0$)
or the cold one (for $B_{0} \le 0$ and $B_{1} < 0$); when compared to the ice
core data this situation is closer to the time interval between 15000 and
27000 years before present, since during that time the system was preferably
in its cold state and the forcing apparently crossed the threshold only
infrequently and during short periods of time.
\begin{table}[t]
\caption{Parameters of the conceptual model. All parameters have the same
  values as in the publication of \citet{Braun2005}. $A_{0}$, $A_{1}$, $B_{0}$ and $B_{1}$ are 
given in freshwater units (i.e. in mSv = milli-Sverdrup; $1~\unit{mSv} = 10^{3}~\unit{m^{3}/s}$), since the 
conceptual model was originally designed to mimic the response of the THC to
an anomaly in 
the surface freshwater flux.}
  \vskip4mm
  \begin{center}
  \label{tab:table1}
  \iftwocol{\small}{}
  \begin{tabular}{cll}
    \tophline
    Parameter & Chosen value \\
    \middlehline
    A$_{0}$ & -27 \unit{mSv}    \\
    A$_{1}$ & 27 \unit{mSv}     \\
    B$_{0}$ & -9.7 \unit{mSv}   \\
    B$_{1}$ & 11.2 \unit{mSv}   \\
    $\tau_{0}$ & 1200 \unit{years} \\ 
    $\tau_{1}$ & 800 \unit{years}  \\
    \bottomhline
  \end{tabular}
  \end{center}
\end{table}

\subsection{Comparison with a coupled climate model}

In order to test our conceptual model we compare its performance under a
number of systematic forcing scenarios with the performance of the far more
comprehensive model CLIMBER-2 (a short description of that model is given in
the Appendix; a detailed description exists in the publication of
\citet{Petoukhov2000}).  

Analogous to \citet{Braun2005}, we investigate the response of both models to
a forcing that consists of two century-scale sinusoidal cycles. In the
conceptual model, the forcing is implemented as the forcing function f. In the
EMIC, the forcing is added to the surface freshwater flux in the latitudinal
belt 50-70 $^{\circ}$N, following \citet{Ganopolski2001} and
\citet{Braun2005}. This anomaly changes the vertical density gradient in the
ocean and can thus trigger DO events. Switches from the cold state into the
warm one are excited by sufficiently large (order of magnitude: a few
centimetre per year in the surface freshwater flux into the relevant area of
the North Atlantic) negative freshwater anomalies (i.e. by positive surface
density anomalies that are strong enough to trigger buoyancy [deep]
convection), and the opposite switches are triggered by sufficiently large
positive freshwater anomalies (i.e. by negative surface density anomalies that
are strong enough to stop buoyancy [deep] convection). This justifies our
choice for the logical relations that govern the dynamics of the transitions
in the conceptual model (i.e. $f(t) < T(t)$ as the condition for the switch
from the cold state to the warm one, $f(t) > T(t)$ for the opposite switch). 

A detailed comparison between both model outputs is presented in the
Supplementary Material. 
We here only summarise the main results: We find a general agreement between both 
models, which is robust when the forcing parameters are varied over
some range 
(Supplementary Figs. 1-6). The conceptual model reproduces the existence of three
different regimes ({\it{cold}}, {\it{warm}}, {\it{oscillatory}}) in the output of the EMIC
and also their approximate position in the
forcing parameter-space. By construction only the 
nonlinear component in the response of the EMIC to the forcing is reproduced by the 
conceptual model (this component represents the saw-tooth shape of DO
events). A second, 
more linear component is not included in the conceptual model (this component represents 
small-amplitude temperature anomalies which are superimposed on the saw-tooth
shaped events 
in the EMIC). In particular, the conceptual model very well reproduces the
timing of the onset of DO events in the EMIC. The fact that our conceptual model, despite its 
simplicity, agrees in so many aspects with the much more detailed model
CLIMBER-2 suggests 
that it indeed captures the key features in the dynamics of DO events in that
model. 

We would like to stress that the output of the EMIC indeed supports our
assumption of an overshooting in the stability of the system during the
transitions between both climate states: When driven by a periodic forcing
(with a period of 1470 years), the EMIC can show periodic oscillations during
which it remains in either of its states for more than one forcing period
(i.e. for considerably more than 1470 years, compare Supplementary Fig.
4a). This implies that (at least in the EMIC) the conditions for a return to
the opposite state indeed ameliorate with increased duration of the cold or
warm intervals. If the thresholds in the model were constant (or gradually
increasing with increasing duration of the simulated cold / warm intervals),
in contrast, the duration of the cold and warm intervals during the simulated
oscillation could never be longer than 1470 years: If a periodic forcing does
not cross a constant (or gradually increasing) threshold within its first
period, it never crosses the threshold, due to the periodicity of the forcing.

\section{Nonlinear resonance mechanisms in the model}
\label{sec:resonance}

Strongly nonlinear systems can show complex and apparently counterintuitive
resonance phenomena that cannot occur in simple linear systems. In this
section we use our conceptual model to demonstrate and to discuss two of these
phenomena, i.e. stochastic resonance (SR) and ghost resonance (GR). Since the
explanation of the 1470-year cycle (and in fact even its significance) is
still an open question, we further discuss how future tests could distinguish
between the proposed mechanisms. 

\subsection{Stochastic resonance (SR)}

\begin{figure}[t]
\vspace*{2mm}
  % don't type the extension (eps or pdf) of the graphics file here
  \begin{center}
  \includegraphics[width=8.5cm]{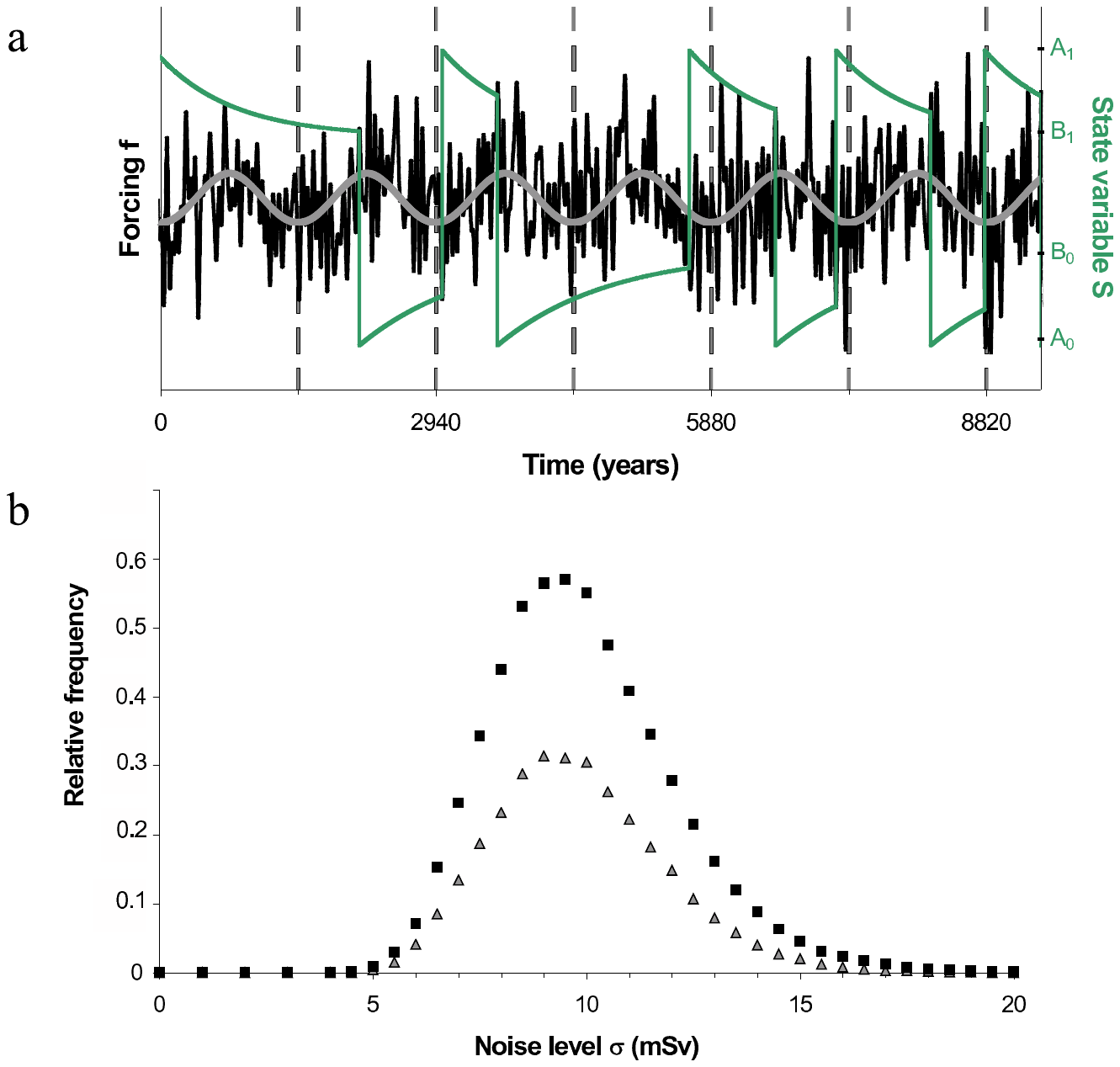}
  \end{center}
  \caption{\label{fig:figure3}
Stochastic resonance. The input consists of: 1. a
sub-threshold sinusoidal signal with a period of 1470 years and an amplitude
of 4.5 mSv (about 40 percent of the threshold value B$_{1}$ above which
DO events occur in the model), 2. a random Gaussian-distributed signal with
white noise power signature (standard deviation $\sigma$ = 8 mSv) and a cutoff
frequency of 1/(50 years). The cutoff is used since no damping exists in the
model and it thus shows an unrealistically large sensitivity to high-frequency
(i.e. decadal-scale or faster) forcing. A: Total input (black), periodic
input component (grey), model output (green). Dashed lines are spaced by 1470 
years. B: Relative frequency to obtain a spacing of 1470 years $\pm$10\% (triangles)
respectively $\pm$20\% (squares) between successive events, as a function of the noise level $\sigma$.   
}
\end{figure}

In linear systems that are driven by a periodic input, the existence of noise 
generally reduces the regularity of the output (e.g. the coherence between the
input and the output). This is not necessarily the case in nonlinear systems:
Excitable or bistable nonlinear systems with a threshold and with noise,
which are driven by a sinusoidal sub-threshold input, can show maximum
coherence between the input and the output for an intermediate noise level,
for which the leading output frequency of the system is close to the input
frequency. This phenomenon is called stochastic resonance (SR)
\citep{Benzi1982, Gammaitoni1998}. SR has been suggested to explain the 
characteristic timing of DO events \citep{Alley2001a}, i.e. the apparent
tendency of the events to occur with a spacing of about 1470 years or integer
multiples thereof. It has further been demonstrated that DO events in the
model CLIMBER-2 can be subject to SR \citep{Ganopolski2002}. 

\begin{figure*}[t]
\vspace*{2mm}
  % don't type the extension (eps or pdf) of the graphics file here
  \begin{center}
  \includegraphics[width=14cm]{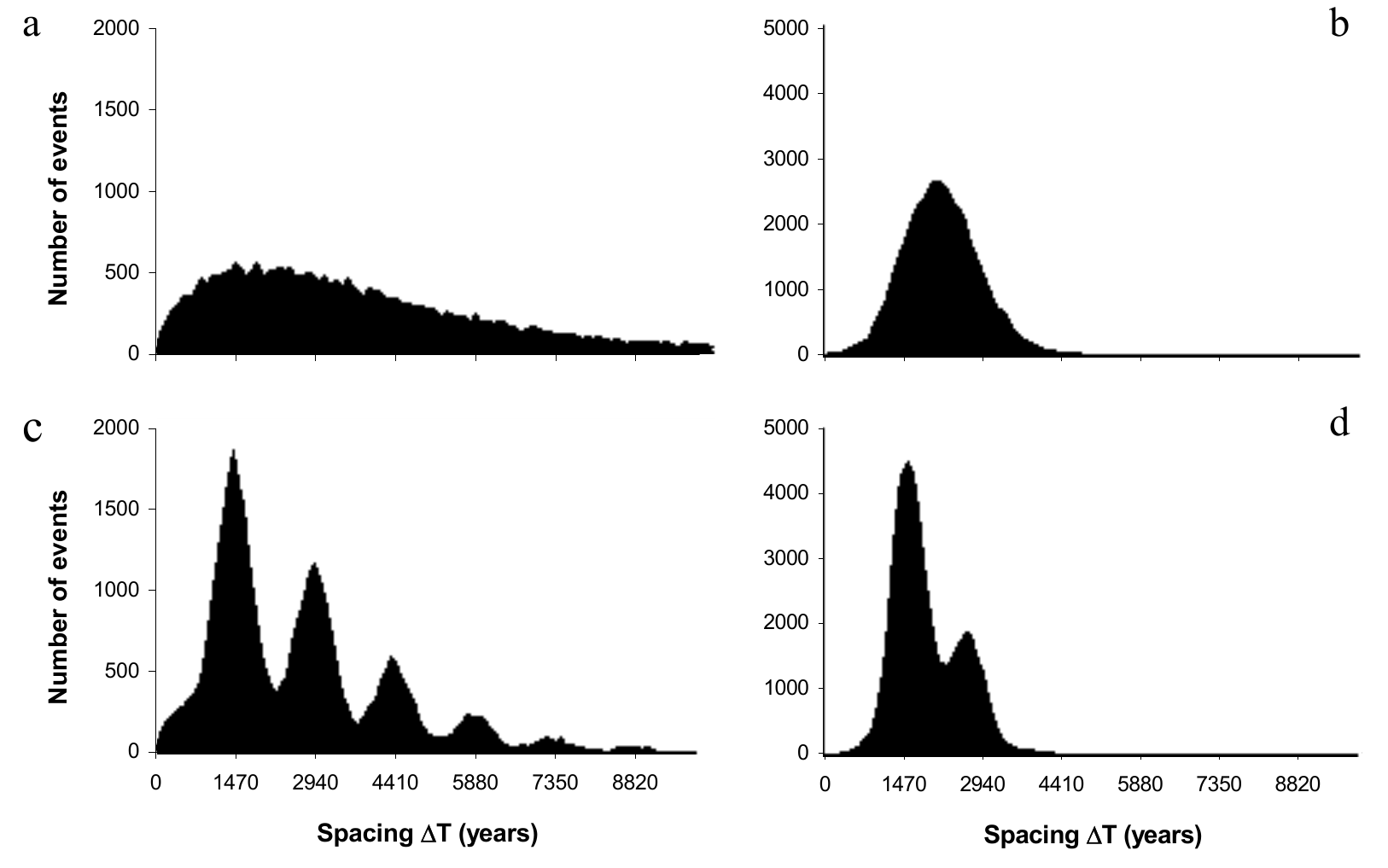}
  \end{center}
  \caption{\label{fig:figure4}
Distribution of the spacing $\Delta$T between successive events. The input in
a and b consists of noise only, with a standard deviation $\sigma$ of 8 mSv
(as in Fig.~\ref{fig:figure3}a). In c and d, a sinusoidal input component
(amplitude = 4.5 mSv, period = 1470 years; compare Fig.~\ref{fig:figure3}) is
added to the noise. In a and c the threshold function is constant in each
state (20 mSv in the warm state and -20 mSv in the cold one), while the
overshooting relaxation assumption is used in b and d (with threshold
parameters as shown in Table 1). Thus, c corresponds to the ``usual'' SR while
d shows our ``overshooting'' SR. 
}
\end{figure*}

Here we apply our conceptual model to reproduce these results and to reanalyse
the underlying mechanism. We use an input that is composed of: 
(i) a sinusoidal signal with a
period of 1470 years, (ii) additional white noise. Figures~\ref{fig:figure3}
and~\ref{fig:figure4} show that for a suitable noise level the model can indeed show DO
events with a preferred spacing of about 1470 years or integer multiples
thereof. The reason for this pattern in the output is easily understandable in
the context of the model dynamics: DO events in the model are triggered by
pronounced minima of the total input (total input = periodic signal plus
noise). These minima generally cluster around the minima of the sinusoidal
signal, and the start of the simulated events thus has a tendency to coincide
with minima of the sinusoidal signal (Fig.~\ref{fig:figure3}a). Some minima of
the sinusoidal signal, however, are not able to trigger an event, because the
magnitude of the noise around these minima is too small so that the threshold
function is not reached by the total input. Consequently, a cycle is sometimes
missed, and the spacing of successive events can change from about 1470
years to multiples thereof.  

Unlike the model CLIMBER-2 (which has a complex relationship between the input
and the output and also a large computational cost) our conceptual model can
be used for a detailed investigation of the SR, e.g. because the
dynamics of the model is simple and precisely known and because probability
measures (such as waiting time distributions) can be explicitly computed. In fact, the 
resonant pattern in the conceptual model (Fig.~\ref{fig:figure3}) is due to
two time-scale matching conditions: The noise level is such that the average
waiting time between successive noise-induced transitions is comparable to
{\it{half}} of the period of the periodic forcing, and also comparable to the
relaxation times $\tau_{0}$ and $\tau_{1}$ of the threshold function (compare
Fig.~\ref{fig:figure4}b). This situation is different from 
the usual SR, in which thresholds (or potentials) are constant in time (apart
from the influence of the periodic input). In the
usual SR, only one time-scale matching condition exists
\citep{Gammaitoni1998}, namely the one that the average waiting time between
successive noise-induced transitions (i.e. the inverse of the so-called
Kramers rate) is comparable to {\it{half}} of the period of the periodic forcing. 

In order to investigate the implications of the second condition we simulate
histograms for four different scenarios in the conceptual model
(Fig.~\ref{fig:figure4}): 1. noise-only input, constant threshold
(Fig.~\ref{fig:figure4}a); 2. noise-only input, overshooting threshold
(Fig.~\ref{fig:figure4}b); 3. noise plus periodic input, constant threshold
(Fig.~\ref{fig:figure4}c); 4. noise plus periodic input, overshooting
threshold (Fig.~\ref{fig:figure4}d). We note that 3. corresponds to the usual
SR, while 4. describes our {\it{overshooting stochastic resonance}}. As can be
seen from the histograms, the existence of the millennial-scale relaxation
process leads to a synchronisation in the sense that the waiting times between
successive events are confined within a much smaller time interval (about
1000-4500 years with the overshooting, compared to about 0-10000 years without
overshooting).   

This confinement is plausible since the transition probability
between both model states strongly depends on the magnitude of the threshold,
which declines with increasing duration of the cold or warm
intervals: When the 
standard deviation of the noise level is chosen such that
the average waiting time between successive noise-induced transitions is
comparable to the relaxation times $\tau_{0}$ and $\tau_{1}$, as in
Fig.~\ref{fig:figure4}, the overshooting relaxation strongly reduces the
transition probability for waiting times much smaller than the relaxation time
(since the corresponding values of the threshold function are large) and
increases the transition probability for waiting times of the order of the
relaxation time or larger (since the corresponding values of the threshold function are
considerably smaller). The probability to find an only century-scale spacing between
successive events is thus small, because the corresponding transition
probabilities are small. On the other hand, the probability to find a
multi-millennial spacing is also small, because the states are already
depopulated before (i.e. the probability to obtain lifetimes considerably
larger than the relaxation time is almost zero). This explains why the
possible values for the spacing between successive DO events are restricted to
a much smaller range than in the usual SR (i.e. in the case with constant
thresholds).  

This synchronisation effect is indeed not unique to the conceptual model: The
output of the coupled model CLIMBER-2 shows a similar pattern (with
possible waiting times between successive DO events of e.g. about 1500-5000 years
or about 1000-3000 years, depending on the noise level; compare Fig. 4a-d in
the publication of \citet{Ganopolski2002}). This similarity is of course not
surprising, since the conceptual model is apparently able to mimic the
events in the EMIC and since an overshooting in the stability of both states clearly also
exists in CLIMBER-2 (compare Sect. 3.3). We note that in the GISP2 ice core data, DO
events in the time interval 27000-45000 years before present (which, as
discussed in Sect. 3.2, is the best analogue to the ``background climate
state'' in our conceptional model, since the duration of the cold and warm
intervals in the data is comparable in that interval) have spacings of about
1000-3000 years (compare Fig. 1) and were reported to cluster around values of
either about 1470 years or about 2940 years \citep{Schulz2002}. Because the SR
mechanism could explain such a pattern (compare Fig.~\ref{fig:figure4}d) it
has originally been proposed. However, this mechanism requires a sinusoidal
input with a period of about 1470 years, which has so far not been detected.

\subsection{Ghost resonance (GR)}
In linear systems which are driven by a periodic input, the frequencies of the
output are always identical to the input frequencies. This is not necessarily
the case in nonlinear systems. For example, nonlinear excitable (or bistable)
systems that are driven by an input with  frequencies corresponding to
harmonics of a fundamental frequency (which itself is not present in the
input) can show a resonance at the fundamental frequency, i.e. at a frequency
with zero input power. This phenomenon, which was first described in order to
explain the pitch of complex sounds \citep{Chialvo2002, Chialvo2003} and later
observed experimentally e.g. in laser systems \citep{Buldu2003}, is called
ghost resonance (GR). GR and SR can indeed occur in the same class of systems,
e.g. in bistable or excitable systems with thresholds. However, unlike SR, GR
requires a periodic driver with more than one frequency. Although many
geophysical systems might be subject to GR (since the relevant processes often
have thresholds), the occurrence of this mechanism has so far not expressly
been demonstrated in geoscience. 

\begin{figure}[t]
\vspace*{2mm}
  % don't type the extension (eps or pdf) of the graphics file here
  \begin{center}
  \includegraphics[width=8.5cm]{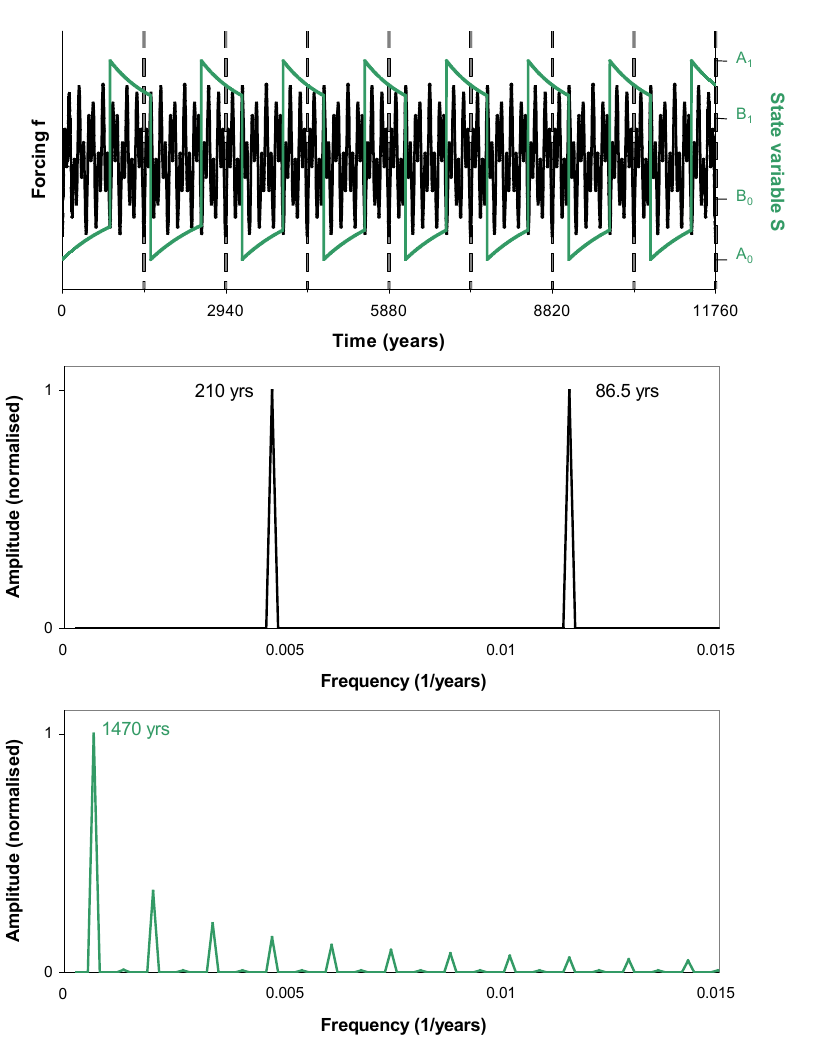}
  \end{center}
  \caption{\label{fig:figure5}
Ghost resonance. Top: Forcing (black) and model response (green). Middle:
Amplitude spectrum of the forcing. Bottom: Amplitude spectrum of the model response. We use two sinusoidal forcing
cycles, with frequencies of 7/(1470 years) and 17/(1470 years), respectively, and with equal 
amplitudes. These two cycles coincide every 1470 years and create peaks
of particularly pronounced magnitude, spaced by exactly that period. Thus,
despite the fact that there is no spectral power at the corresponding
frequency (see middle panel), the forcing repeatedly crosses the threshold at
those intervals. Consequently, the response of the conceptual model (i.e. 
the time evolution of the state variable S) shows strictly repetitive DO events with 
a period of 1470 years (as indicated by the dashed lines, which are spaced by 
1470 years). Despite the lack of a 1470-year spectral component in the forcing, the 
output shows a very prominent peak at the corresponding frequency.
}
\end{figure}

Here we discuss a hypothesis that was recently proposed to explain the
1/(1470 years) leading frequency of DO events \citep{Braun2005}. The underlying mechanism of
the hypothesis is in fact the first reported manifestation of GR in a
geophysical model system. According to that hypothesis the 1/(1470 years)
frequency of DO events could represent the nonlinear climate response
to forcing cycles with frequencies close to harmonics of 1/(1470 years). Our
conceptual model illustrates the plausibility of this mechanism: We use a 
bi-sinusoidal input with frequencies of 7/(1470 years) and 17/(1470 years), i.e. with 
frequencies corresponding to the 7th and the 17th harmonic of a 1/(1470 years) 
fundamental frequency, and with equal amplitudes. A spectral component corresponding 
to the fundamental frequency is not explicitly present in the input. Since the two 
sinusoidal cycles correspond to harmonics of the missing fundamental, the input signal 
repeats with a period of 1470 years. For an appropriate range of input amplitudes, the 
output of the conceptual models shows periodic DO events with a period of 1470 years 
(Fig.~\ref{fig:figure5}). Unlike the input, the model output exhibits a
pronounced frequency of 1/(1470 years), corresponding to the leading frequency
of DO events and to the fundamental frequency that is absent in the
input. This apparent paradox is explained by the fact that the two driving
cycles enter in phase every 1470 years, thus creating pronounced peaks spaced
by that period. Because the magnitude of these peaks results from constructive
interference of the two driving cycles, it is indeed robust that a threshold
process can be much more sensitive to these peaks than to the two original driving
cycles \citep{Chialvo2003}.
   
The main strength of the GR mechanism is that -- unlike the SR mechanism -- it
can relate the leading frequency of DO events to a main driver of natural
(non-anthropogenic) climate variability, since proxies of solar activity
suggest the existence of solar cycles with periods close to 1470/7 (=210)
years (De Vries or Suess cycle) and 1470/17 ($\approx$86.5) years (Gleissberg
cycle) \citep{Stuiver1993, Wagner2001, Peristykh2003}. So far, however, no
empirical evidence for this mechanism has been found \citep{Muscheler2006},
nor has it been shown yet that changes in solar activity over the solar cycles
are sufficiently strong to actually trigger DO events.  

\begin{figure}[t]
\vspace*{2mm}
 % don't type the extension (eps or pdf) of the graphics file here
\begin{center}
\includegraphics[width=8cm]{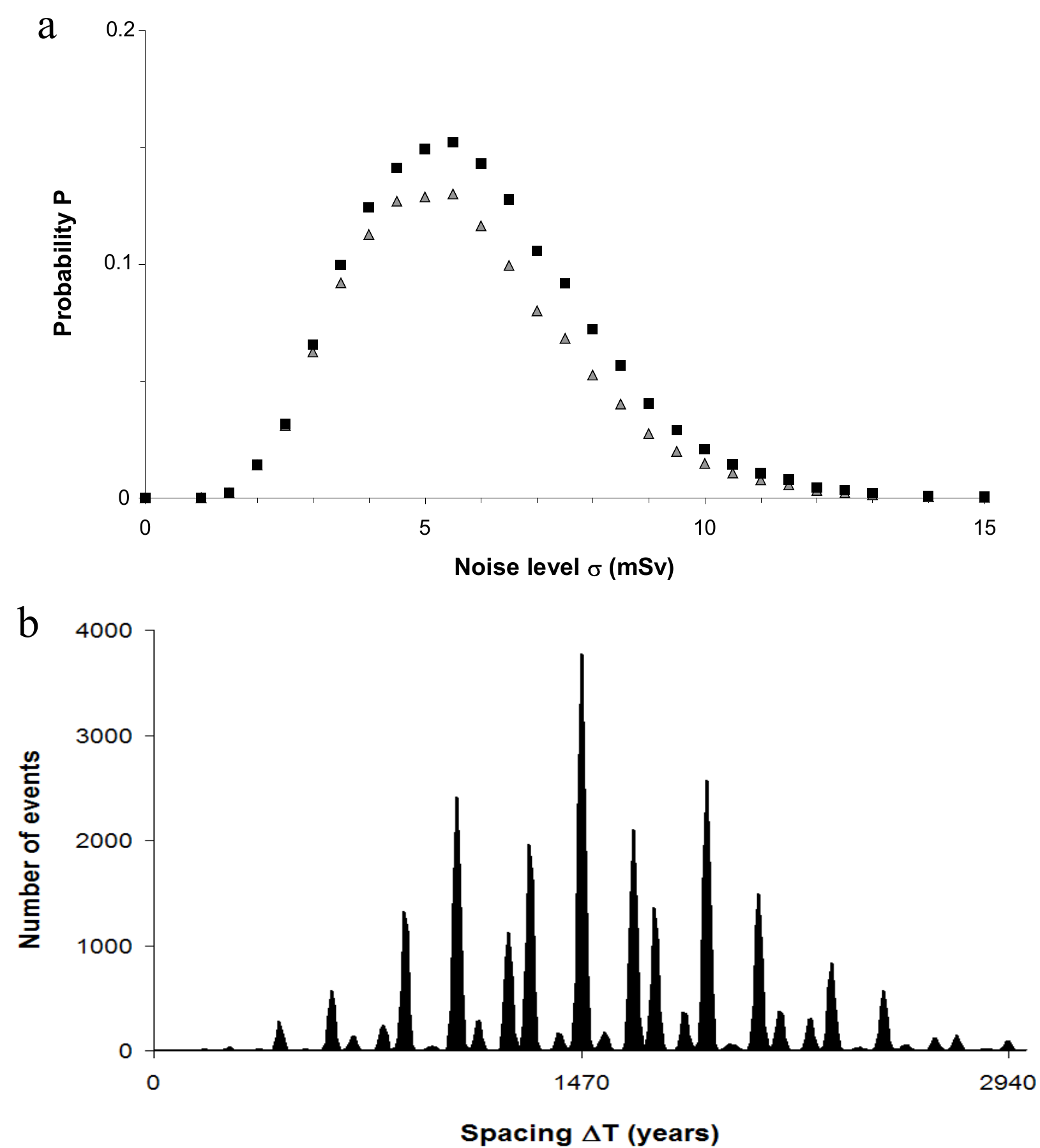}
\end{center}
\caption{\label{fig:figure6}
Ghost stochastic resonance. The input consists of: 1. two sinusoidal forcing
cycles with frequencies of 7/(1470 years) and 17/(1470 years), respectively,
and with an amplitude of 8 mSv, 2. a random Gaussian-distributed signal with
white noise power signature and a cutoff frequency of 1/(50 years), as in
Fig.~\ref{fig:figure3}. In a, the relative frequency to obtain a spacing
of 1470 years $\pm$1\% (triangles) respectively $\pm$2\% (squares) between
successive events is shown as a function of the noise level $\sigma$. B shows
the distribution of the spacing $\Delta$T between successive events (standard
deviation of the noise: 5.5 mSv).
}
\end{figure}

In order to investigate the stability of this mechanism we further add a
stochastic component (i.e. white noise) to the forcing. In this case the
events are -- of course -- not strictly periodic anymore. Similar to the SR
case, an optimal (i.e. intermediate) noise level exists for which the waiting
time distribution of the simulated events exhibits a maximum at a value of
1470 years, corresponding to the period of the fundamental frequency of the
two input cycles (Fig.~\ref{fig:figure6}). In contrast to the SR case, in which a fairly
simple waiting time distribution with a few broad maxima of century scale
width is obtained (compare Fig.~\ref{fig:figure4}d), we now find a much more
complex pattern with a large number of very sharp lines of only decadal scale
width. Since the waiting time distributions of both mechanisms are
considerably different, it could -- at least in principle -- be possible to
distinguish between both mechanisms by analysing the distribution of the
observed DO events. In practise, however, this approach is complicated by the
fact that only about ten events appear to be sufficiently well dated
for this kind of analysis, and even their spacing has an uncertainty of about
50 years \citep{Ditlevsen2007}, which is already of the same order as the
width of the peaks in Fig.~\ref{fig:figure6}b. We note that the mechanism that
is described in Fig.~\ref{fig:figure6} is known as ghost stochastic resonance (GSR), and its
occurrence and robustness has already been reported before in other systems
with thresholds and multiple states of operation \citep{Chialvo2003}. At least
in our system, however, this mechanism is even more complex than the other two types of
resonance (SR, GR). It is beyond the scope of this paper to describe the GSR
mechanism in more detail.  

\subsection{Testing the proposed mechanisms}

The most direct way to test which of the proposed mechanisms -- if any
-- provides the correct explanation for the timing of DO events would be to
reconstruct decadal-scale density anomalies in the North Atlantic in
connection with the events. This is not possible since even the most highly
resolved oceanic records do not allow to reconstruct the variability of the
glacial ocean on that time scale. Thus, only indirect tests can be
performed. The identification of the postulated 1/(1470 years) forcing
frequency, which has so far not been detected, would certainly give further
support for the SR mechanism. And in order to support the GR mechanism, it
remains crucial to demonstrate that century-scale solar irradiance variations
are indeed of sufficiently large amplitude to trigger repeated transitions
(with a preferred time scale of about 1470 years) between the two glacial
climate states. This could be tested with climate models.

An elegant and simple test is to make use of the observation that DO
events in the Earth system model CLIMBER-2 represent the nonlinear response to
the forcing, and that an additional (and much smaller) linear response is
superimposed on the events. In the absence of any threshold crossing (e.g. in
the Holocene, during which DO events did not occur) the response to the
forcing, in contrast, does not show a strong nonlinear component. This
suggests that Holocene climate archives from the North Atlantic region might
be able to reveal what triggered glacial DO events. This approach has two
major advantages: First, more reliable (e.g. better resolved and dated)
records are available to solve this issue. Second, linear analysis methods can
be used for that purpose, e.g. linear correlations. In the context of the GR
mechanism, for example, the existence of a pronounced correlation between
Holocene climate indices from the North Atlantic and solar activity proxies
(reconstructed e.g. from $^{14}$C variations in precisely dated tree rings)
would be expected. Up to now at least one study exists that supports this
prediction of a linear relationship between century-scale solar forcing and
North Atlantic climate variability throughout the Holocene: Proxies of drift
ice anomalies in the North Atlantic show a persistent correlation and a
statistically significant coherency with ``rapid (100- to 200-year),
conspicuously large-amplitude variations'' in solar activity proxies
\citep{Bond2001}, in accordance with the proposed GR mechanism.   

The most challenging test, however, is the direct analysis of the glacial climate records. We are
convinced that one of the main difficulties in this approach is the high
degree of nonlinearity of the events, which -- according to our interpretation
-- has so far not been adequately addressed in many previous studies. For example,
several attempts have already been made in order to investigate the 1470-year
cycle by means of linear spectral analysis methods, and significance levels
have commonly been calculated by assuming a red noise background. To us this
assumption seems to be oversimplified, since the system responds at a
preferred time scale even when driven by white noise (compare
Fig.~\ref{fig:figure4}b). We thus suspect that the significance levels obtained
by this method are unrealistically high. We further think that the
reported lack of a clear phase relation between solar activity proxies and DO
events \citep{Muscheler2006} cannot rule out the idea that solar forcing
synchronised DO events, since in the case of an additional stochastic
forcing component (i.e. in the GSR case) the events are triggered by the
{\it{combined}} effect of solar forcing and noise. Thus, the observed lack could also
imply that only some of the events were in first place triggered by the Sun,
whereas others were caused mainly by random variability (e.g. by noise).

A new and promising approach, which is based on a Monte-Carlo method, has
recently been proposed in order to test the significance of the glacial 1470-year
climate cycle: \citet{Ditlevsen2007} define a certain measure in order to distinguish
between different hypotheses for the timing of DO events, and they explicitly
calculate the value of this measure for the series of events observed in the
ice core. They then compare the calculated value with the values obtained by
several hypothetic processes, e.g. by a random process (for which
assumptions concerning the probability distribution of the recurrence times of
the events have to be made). Significance levels are obtained from the
(numerically estimated) probability distributions of the measure as generated
by the considered process. Although we do not share their conclusions (because
we think that more adequate measures can be chosen, which give considerably
different results) we think that this approach is elegant because significance
levels are not calculated based on linear theories. The method is thus also
applicable to highly nonlinear time series. A major hurdle in this method is that for each considered
process the probability distribution of the waiting times -- which is unknown
for almost all processes -- somehow has to be specified. For example,
Ditlevsen et al. use a simple mathematical (i.e. an exponential) distribution
in order to mimic random DO events. In order to improve their novel
approach, some method is thus needed to calculate waiting time distributions
in response to any possible input. Comprehensive models are not applicable,
due to their large computational cost. Our conceptual model, in contrast, is
well designed for that purpose because it is combines the ability to mimic the
complex nonlinearity of DO events as described by an accepted Earth system
model with the extremely low computational cost of a very simple (zeroth
order) model. We thus think that our work is an important step in
order to develop improved statistical analysis methods which are able to cope
with the extreme nonlinearity of DO events.  

\conclusions[Discussion and conclusions]\label{sec:end}
% If you want to give your conclusions section a title other than simply
% "Conclusions", you can add an optional argument to the \conclusions
% command in SQUARE brackets, e.g.: \conclusions[Summary and conclusions]

We here discussed DO events in the framework of a very simple conceptual model
that is based on three key assumptions, namely (i) the existence of two
different climate states, (ii) a threshold process and (iii) an overshooting
in the stability of the system at the start and the end of the events, which
is followed by a millennial-scale relaxation. These assumptions are in
accordance with paleoclimatic records and/or with simulations performed with
CLIMBER-2, a more complex Earth system model. In a couple of systematic tests
we showed (in the Supplementary Material) that despite its simplicity, our
model very well reproduces DO events as simulated with CLIMBER-2, whose
dynamics is based on a (albeit reduced) description of the underlying
hydro-/thermodynamic processes. The correspondence between both models thus
strengthens our interpretation that the conceptual model can successfully
mimic key features of DO events, and that these can be regarded as a new type
of non-equilibrium oscillation (i.e. as an {\it{overshooting relaxation
    oscillation}}) between two states of a nonlinear system with a threshold.    

Although we discussed our model dynamics in the context of the (thermohaline)
ocean circulation, our model does not explicitly assume that DO events are
linked with changes in the ocean circulation: Threshold behaviour and multiple
states exist in many compartments of the climate system (not only in the
ocean, but e.g. also in the atmosphere and in the cryosphere). Our model thus
cannot rule out a leading role of non-oceanic processes in DO
oscillations. The millennial time scale of the events (which is represented in
our model by the assumption of a millennial-scale relaxation), however,
corresponds to the characteristic time scale of the thermohaline circulation
and thus points to a key role of the ocean in DO oscillations.

The main strength of our model is its simplicity: Due to the obvious relationship 
between forcing and response, the model can demonstrate why even a simple
bistable (or excitable) system with a threshold can respond in a complex way
to a simple forcing, which consists of only one or two sinusoidal inputs and
noise. We applied our model to discuss two highly nonlinear and apparently
counterintuitive resonance mechanisms, namely stochastic resonance and ghost
resonance. In doing so we reported a
new form of stochastic resonance (i.e. an {\it{overshooting stochastic
    resonance}}), in which the overshooting of the system leads to a further
synchronisation effect compared to the usual stochastic resonance. Our
study provides the first explicitly reported manifestation of ghost resonance
in a geophysical (model) system. Since threshold behaviour and multiple
equilibria are not unique to DO events but exist in many geophysical
systems, we would indeed expect that ghost resonances could be inherent in
many geosystems and not just in our model.     

In addition to its applicability to demonstrate and interpret nonlinear
resonance mechanisms, and to test their stability, we further illustrated the
ability of our conceptual model to simulate probability measures (e.g. waiting time distributions, which
are required in order to test the significance and the cause of the proposed
glacial 1470-year climate cycle by means of Monte-Carlo simulations). Because
it combines the ability to reproduce essential aspects of DO
events with the extremely low computational cost of a conceptual model (which
is up to 10$^{7}$ times lower than in the Earth system model CLIMBER-2), we
think that our model represents an important advance in order to develop
adequate nonlinear methods for improved statistical analyses on DO events.

\section*{Appendix: Description of CLIMBER-2}

\normalsize
The Earth system model CLIMBER-2, which we used for our analysis, has 
dynamic components of the atmosphere, of the oceans (including sea 
ice) and the vegetation. Dynamic ice sheets were not included
in our study. CLIMBER-2 is a global model with coarse 
resolution: For the atmosphere and the continents the spatial 
resolution is 10$^{\circ}$ in latitude, and 7 sectors are considered in 
longitude. The ocean is zonally averaged with a latitudinal 
resolution of 2.5$^{\circ}$ for the three large ocean basins. A detailed 
description of the model is given in the publication of \citet{Petoukhov2000}.

DO events in the model represent abrupt switches between two different climate 
states ({\it{stadial}} [i.e. cold] and {\it{interstadial}} [i.e. warm]), 
corresponding to two different modes of the THC: In the interstadial
mode, North Atlantic deep water (NADW) forms at about 65 $^\circ$N and much of
the North Atlantic is ice-free. In the stadial mode, NADW forms
at about 50 $^\circ$N and a considerably larger area of the North Atlantic is
ice-covered. We note that for the climatic background conditions 
of the Last Glacial Maximum (LGM) only the stadial mode is stable in the
model whereas the interstadial mode is excitable but unstable
\citep{Ganopolski2001}. Moreover, the stability of both modes depends on the
actual climate state (e.g. on the configuration of the Laurentide ice sheet
and on the freshwater input into the North Atlantic), and the stability
properties of the system change when the background conditions are modified
(more precisely, the system can be bistable or mono-stable).  

Transitions between both modes can be triggered by anomalies in the 
density field of the North Atlantic, for example by variations in the 
surface freshwater flux (since the density of ocean water increases 
with increasing salinity). In our study we thus implement the forcing 
as a perturbation in the freshwater flux (in the latitudinal belt 
50-70 $^{\circ}$N): We start the model with the climatic background conditions 
of the Last Glacial Maximum (LGM). Following earlier simulations
\citep{Braun2005} we then add a small constant offset of 17 \unit{mSv}
($1~\unit{mSv} = 10^{3}~\unit{m^{3}/s}$) to the  freshwater flux. For this
climate state (which we label perturbed LGM) the THC is in fact bistable
and DO events can be triggered more easily than for LGM conditions. This
perturbed LGM state gives us the background conditions for the model
simulations as presented in this paper.    

\begin{acknowledgements}
The authors thank R. Calov, A. Mangini, S. Rahmstorf, K. Roth and A. Witt for
discussion, and P. Ditlevsen (in particular for observing the difference
between the usual stochastic resonance and our overshooting stochastic resonance) and
two anonymous reviewers for helpful comments. H. Braun was funded by Deutsche
Forschungsgemeinschaft, DFG project number MA 821/33.
\end{acknowledgements}

%% TWO METHODS FOR INCLUDING THE BIBLIOGRAPHY (LIST OF REFERENCES)

%% EITHER TYPE IN THE ENTRIES YOURSELF AS SHOWN HERE IN

%% `thebibliography' ENVIRONMENT,

%%        OR

%% USE THE FOLLOWING TWO COMMANDS SO THAT BIBTEX WILL GENERATE

%% `thebibliography' TEXT FOR YOU AND READ IT IN.

%%

%% \bibliographystyle{egu} %<-- LIST OF REFERENCES TO BE IN "egu.bst" STYLE

%% \bibliography{sample}   %<-- REFERENCES ARE IN FILE "sample.bib"

%%

%%  IF THE ABOVE TWO COMMANDS ARE USED, THEN thebibliography ENVIRONMENT

%%  MUST BE REMOVED.

\end{document}
}